\documentclass[a4paper, 5p, sort&compress, preprint]{elsarticle}

\usepackage{amssymb}
\usepackage{amsthm}
\usepackage{amsmath}
\usepackage{units}
\usepackage{lineno}
\usepackage{isotope}
\usepackage[pdftex, draft=false, debug=false, colorlinks, linkcolor=black, citecolor=black, urlcolor=black, breaklinks=true, hypertexnames=true]{hyperref}
\usepackage{stfloats} 
\usepackage{microtype} 
\usepackage[all]{hypcap} 
\usepackage{booktabs}

\journal{Nuclear Instruments and Methods in Physics Research Section A}

\begin{document}

\begin{frontmatter}

\title{In-situ measurement of the scintillation light attenuation in liquid argon in the \textsc{Gerda} experiment}

\author[tud]{N.~Barros\fnref{presentaddressnuno}}
\fntext[presentaddressnuno]{Present Address: Laboratório de Instrumentação e Física Experimental de Partículas, Lisboa, Portugal}

\author[tud]{A.R.~Domula}
\author[tud]{B.~Lehnert\fnref{presentaddressbjoern}}
\fntext[presentaddressbjoern]{Present Address: Nuclear Science Division, Lawrence Berkeley National Laboratory, Berkeley, California 94720, USA}

\author[tud]{B.~Zatschler\corref{correspondingauthor}}
\cortext[correspondingauthor]{Corresponding author}
\ead{Birgit.Zatschler1@tu-dresden.de}

\author[tud,mta]{K.~Zuber}

\address[tud]{Institute of Nuclear and Particle Physics, TU Dresden, Dresden, Germany}
\address[mta]{MTA Atomki, Institute for Nuclear Research, Hungarian Academy of Sciences, Debrecen, Hungary}

\begin{abstract}
The \textsc{Gerda} experiment searches for the neutrinoless double beta ($0\nu\beta\beta$) decay in $^{76}$Ge in order to probe whether the neutrino is a Majorana particle and to shed light on the neutrino mass ordering. For investigating such a rare decay it is necessary to minimize the background of the experiment.
In Phase~II of the \textsc{Gerda} experiment the scintillation light of liquid argon (LAr) is used as an additional background veto. In order to estimate the efficiency of such a LAr veto it has to be known how far the scintillation light, which peaks at 128\,nm, can travel within the LAr. A dedicated setup was built to measure the attenuation length of the scintillation light in the LAr in-situ within the cryostat of \textsc{Gerda}.
The setup is composed of a stainless steel housing with a photomultiplier tube (PMT) at one side and a moveable $^{90}$Sr source at the other side to measure the light intensity at different distances between source and PMT.

Furthermore, a sophisticated simulation was developed in order to determine the solid angle correction as well as the background for this measurement. 
The analysis results in an absorption length of $15.8 \pm 0.7 (\text{stat}) {}^{+1.5}_{-3.2} (\text{syst})\,\unit{cm}$ under the assumption of a scattering length of 70\,cm at 128\,nm. 
The obtained value of the absorption length is specific for the LAr in \textsc{Gerda} at the time of the measurement.
\end{abstract}


\end{frontmatter}

\section{Introduction}
\label{sec:intro}

The germanium detector array (\textsc{Gerda}) is an experiment located in the underground laboratory Laboratori Nazionali del Gran Sasso (LNGS) and searches for the neutrinoless double beta ($0\nu\beta\beta$) decay in $^{76}$Ge \cite{gerda_p2}. \textsc{Gerda} operates isotopically enriched high purity germanium (HPGe) detectors in liquid argon (LAr), which serves as a cooling and shielding against external radiation.  

The current limit of \textsc{Gerda} on the half live of the $0\nu\beta\beta$ decay of $^{76}$Ge is $T_{1/2} > 0.9 \cdot 10^{26}\,\unit{yr}$ (90\% C.L.) \cite{gerda_limit}. The search for such a rare decay requires to minimize the background of the experiment. 
One of the major improvements in \textsc{Gerda} Phase~II is an additional veto system surrounding the HPGe detector array using the LAr scintillation light.
Excess energy that gets deposited in the LAr will trigger the scintillation light providing a signal to further suppress background.
This includes intrinsic impurities in the LAr that could mimic the double beta decay as well as nuclides produced by cosmogenic activation inside the detectors and other natural abundant nuclides inside the surrounding material like detector holders and cables. 
For this purpose, the LAr cryostat in \textsc{Gerda} was instrumented with a system consisting of a combination of reflectors, wavelength shifters (WLS) and photomultiplier tubes (PMT).

The scintillation light in LAr is created by the emission of a de-excitation photon of 128\,nm from either a singlet or a triplet state of an excited LAr molecule. 
The ratio between the population of singlet and triplet state is 0.3 for electrons.
While the transition of the singlet state is allowed (4 -- 7\,ns), the triplet state is forbidden (1.0 -- 1.7\,$\mu$s), resulting in different decay times \cite{triplet_lifetime}. 

In case the excited LAr molecule collides with impurities, such as oxygen or nitrogen, it can de-excite non-radiatively, decreasing the light yield as well as the lifetime of the triplet state. Furthermore, oxygen can absorb scintillation photons directly, which also decreases the light yield as well as the absorption length even more.

Consequently, the scintillation light cannot propagate infinitely inside LAr but is attenuated due to present impurities\cite{att_meas_3,att_meas_4,att_meas_5}, which causes a limit on the effective active volume of the LAr veto. In order to estimate its efficiency the attenuation length of the scintillation light was measured inside the \textsc{Gerda} cryostat. 

\begin{figure*}[t!]
\centering
\includegraphics[width=18cm]{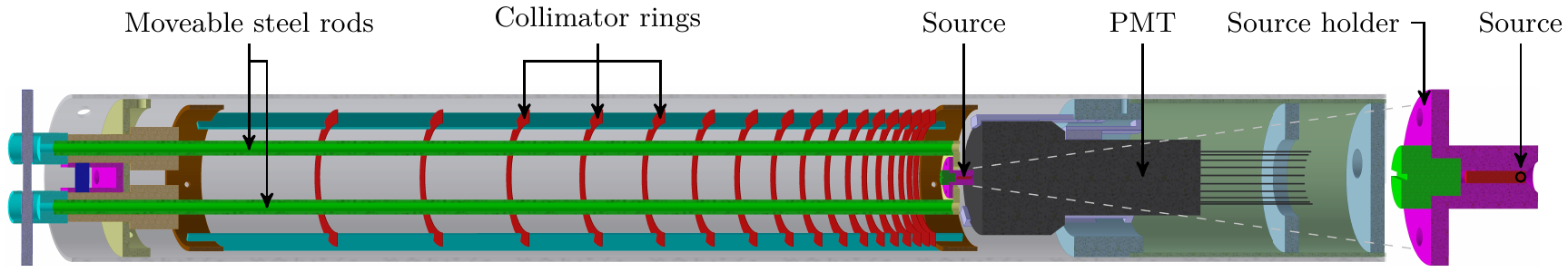}
\caption{CAD drawing of the setup for the attenuation measurement. The setup is completely made of stainless steel with the exception of the PMT holder, which is composed of 
PTFE. 
The PMT is fixed on the right side of the setup, while the source can be moved from the left side to adjust the distance between them.
The source is located in a source holder which is attached to the movable steel rods. The movement is controlled by a stepper motor (not shown in this figure), which is situated at the left end of the setup and connected via cogwheels to the steel rods. 
The whole setup is 1\,m long and the source can be moved at distances from 0.6\,cm to 55.6\,cm with respect to the PMT.
The collimator rings absorb scattered and reflected light from the inner steel wall of the setup. They are arranged in a way that light reflected at the steel wall cannot reach the PMT directly.}
\label{fig:setup}
\end{figure*}

\newpage
Since the attenuation length is just one of the key parameters and other important values are still poorly known, the estimation of the LAr efficiency is investigated by a subgroup of \textsc{Gerda} and is not covered in this work.

An in-situ measurement has been performed because the attenuation is strongly dependent on the specific impurities present in the \textsc{Gerda} LAr cryostat. 
The results of light attenuation measurements of other experiments cannot be applied to \textsc{Gerda} due to their different impurity content.

In the experiment it is not possible to disentangle scattering $\alpha_\text{scat}$ and absorption $\alpha_\text{abs}$ lengths and only the attenuation length $1/ \alpha_\text{att} = 1/ \alpha_\text{abs} + 1/ \alpha_\text{scat}$ is measured. For the implementation of $\alpha_\text{abs}$ and $\alpha_\text{scat}$ into the simulation, $\alpha_\text{abs}$ is determined under different hypotheses of $\alpha_\text{scat}$. This is done because $\alpha_\text{scat}$ is less dependent on impurities in LAr, there are better estimates of it and assuming typical values, it is not the dominant effect in \textsc{Gerda}.

\section{Experimental setup}
\label{sec:setup}

For the measurement of the attenuation of the scintillation light in the LAr inside \textsc{Gerda} a dedicated setup was designed that could be submerged directly into the LAr cryostat. A CAD drawing of the setup is shown in figure~\ref{fig:setup}. 

The setup consists of a steel housing with a PMT mounted on one end and a 7\,kBq \isotope[90]{Sr} source in front of the PMT, which is held by movable steel rods.
The total length of the setup is 1\,m and the source can be moved by the steel rods in order to measure the scintillation light at a distance from 0.6\,cm to 55.6\,cm between PMT and source. The measurement took place at 12 various distances with different step lengths as shown in figure~\ref{fig:FADC_spectra}.

\begin{figure}[ht!]
\centering
\includegraphics[width=8.8cm]{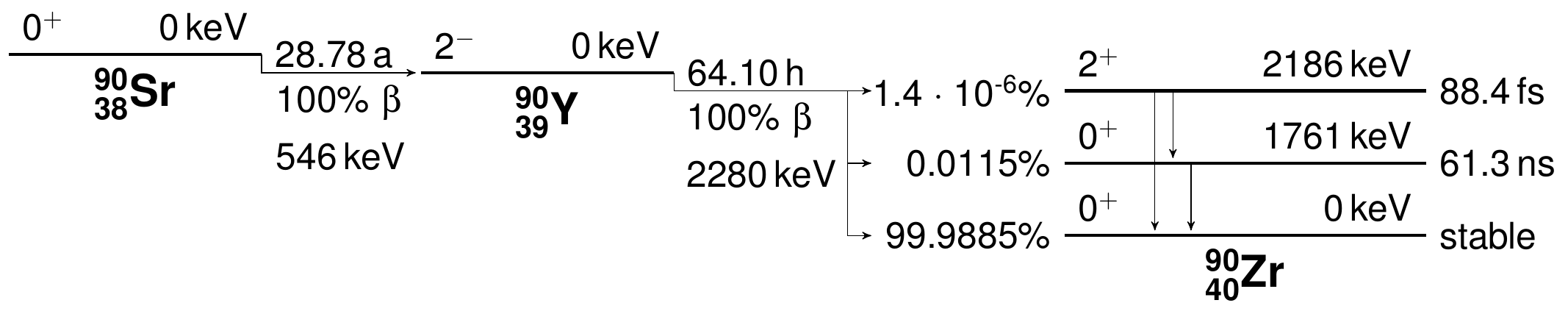}
\caption{Decay scheme of \isotope[90]{Sr}. Values are taken from \cite{toi}.}
\label{fig:Sr90_decay_chain}
\end{figure}

The $\beta$ particles emitted by the source deposit energy in the LAr and trigger scintillation light which can be read out by the PMT. As shown in figure~\ref{fig:Sr90_decay_chain}, \isotope[90]{Sr} is an almost pure $\beta$ emitter, which is essential for this measurement, because gamma radiation can travel very far in LAr, while beta particles create a nearly point like source of scintillation light. In order to minimize the uncertainty of the distance traveled by the scintillation photons to the PMT, their creation needs to be as point-like as possible.
 
Since the scintillation light is attenuated during its propagation through LAr, the measured $\beta$ spectrum of the \isotope[90]{Sr} source is shifted to lower energies at longer distances as shown in figure~\ref{fig:FADC_spectra}. Obviously, the larger the distance between source and PMT, the less light reaches the PMT.

\begin{figure}[ht!]
\centering
\includegraphics[width=8.8cm]{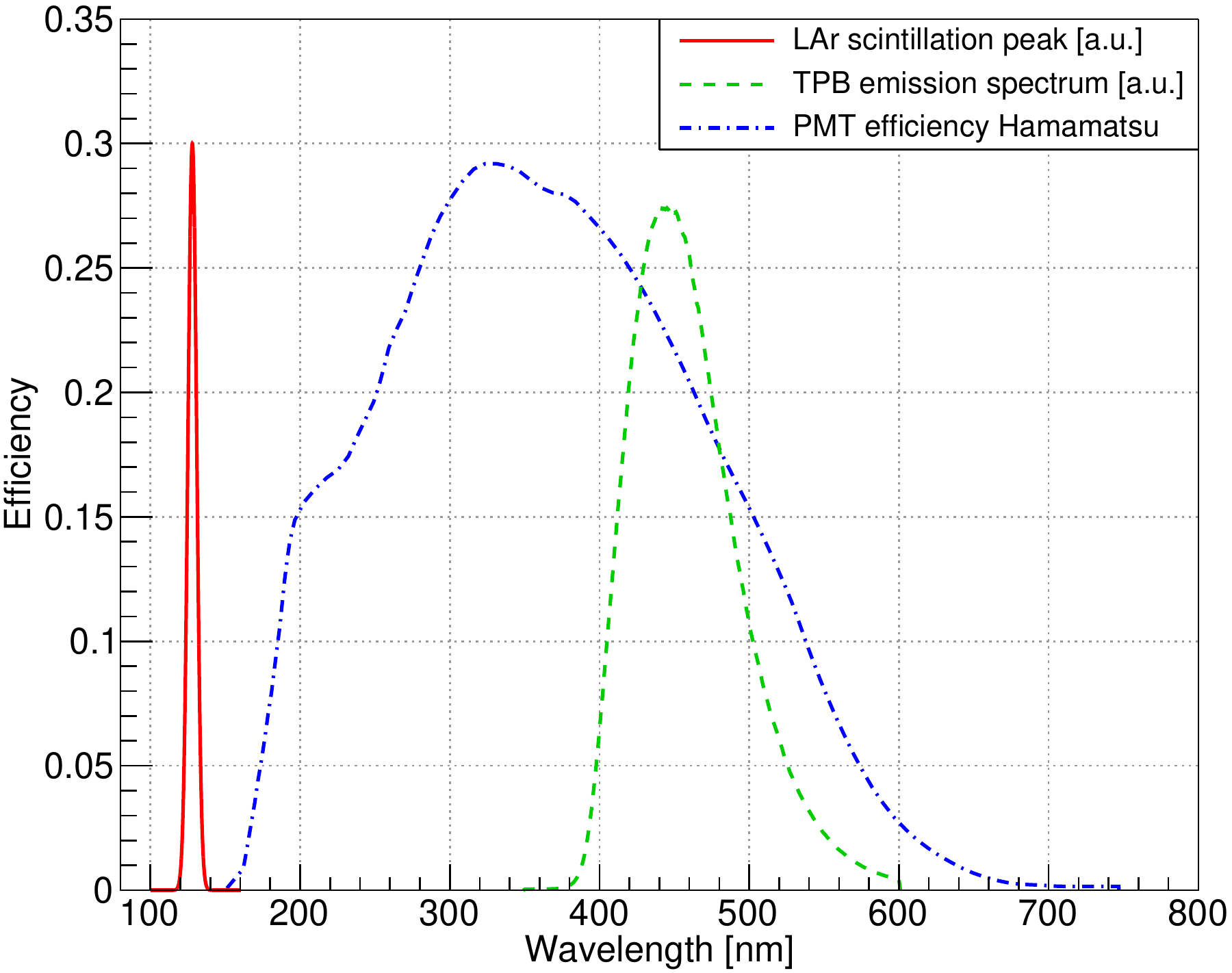}
\caption{The LAr scintillation light is emitted in a sharp peak around 128\,nm \cite{LAr_spectrum} and shifted by TPB \cite{TPB_emission} to the sensitive region of the PMT \cite{Hamamatsu}.}
\label{fig:PMT_eff+TPB}
\end{figure}

A series of stainless steel rings were mounted inside the setup to collimate the scintillation light, blocking photons reflected from the inner steel wall. 
A stepper motor is connected via cogwheels to the steel rods to move the source. All parts of the setup are made of stainless steel, except for the PMT holder, which is made of polytetrafluoroethylene (PTFE). 
A \href{http://lartpc-docdb.fnal.gov/0004/000441/001/R11065_data_sheet_0903-1.pdf}{R11065-10} 3" PMT from Hamamatsu was chosen due to its low intrinsic radioactivity and its particular design for cryogenic operation. The PMT is coated with the WLS tetraphenyl butadiene (TPB), in order to shift the scintillation photons from the ultra-violet regime to the region where the PMT is most sensitive (the peak of the TPB emission spectrum is at 445\,nm). Figure~\ref{fig:PMT_eff+TPB} shows the LAr scintillation peak, the emission spectrum of TPB and the efficiency curve of the PMT.

\section{Data taking and signal reconstruction}
\label{sec:data}

Pulse traces of the PMT were recorded with a 14-bit Fast Analog to Digital Converter (FADC) board (\href{http://www.struck.de/sis3301.htm}{Struck, SIS3301}) with a sampling frequency of 100\,MHz and a trace length of 131072 samples (approx.\ 1.3 ms). 

All traces were saved for each source position to allow for a complex offline analysis. In a first step, all these traces were cleaned from a periodic interference of 20 kHz caused by the operation of the stepper motor.

\newpage
The motor interference would falsify the baseline determination before and after an event as well as the integral of the event. 
That is why these interferences are flagged in the data cleaning in order to reject events overlaid with such a distortion.

\begin{figure*}[b!]
\centering
\includegraphics[width=16cm]{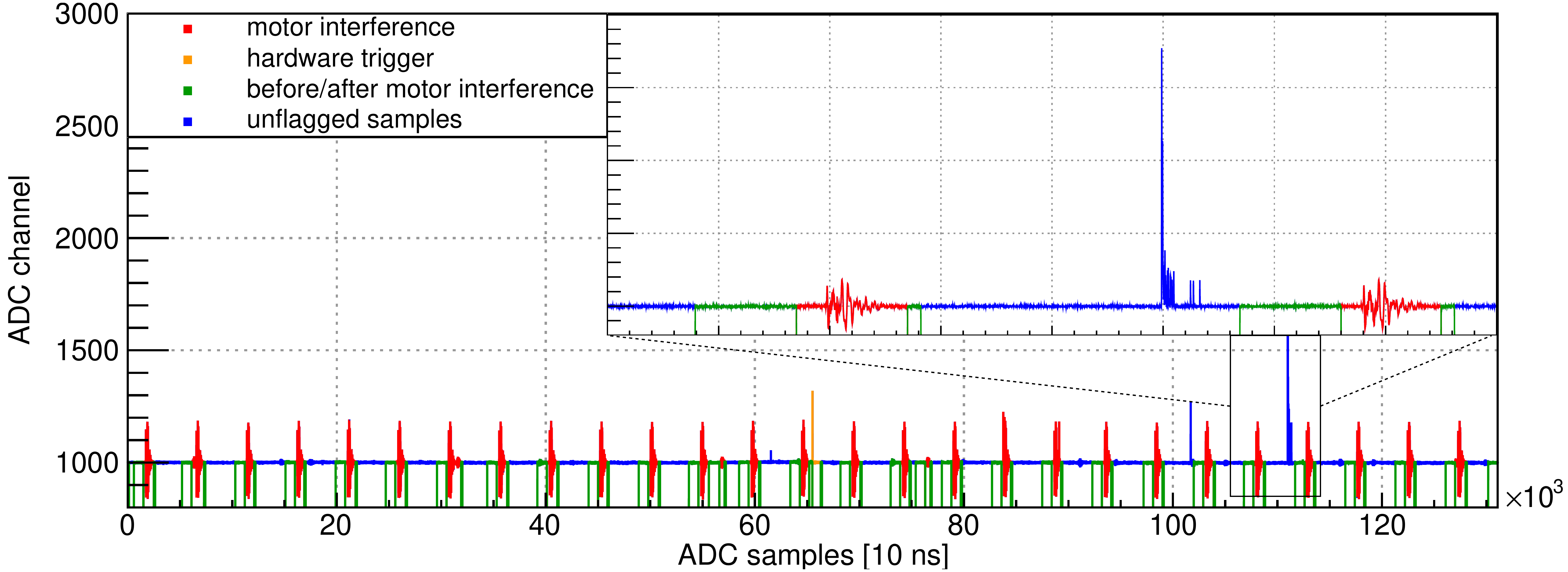}
\caption{(Color online)
Pulse trace with flags indicating the motor interferences. 
The hardware trigger is flagged in order to avoid a systematic bias of larger events.
Only the blue samples are considered for finding physical events. The green samples are safe to be used for the event reconstruction and baseline determination. Physical events can start in the blue and finish in the blue or green area as long as there are enough samples available for the baseline determination before and after the event.}
\label{fig:trace_flagged}
\end{figure*}

An example trace can be seen in figure~\ref{fig:trace_flagged}.
The hardware trigger used for the decision whether to store a pulse trace is located exactly in the middle of every pulse trace.

Events created by only a few photons cannot exceed the condition of the hardware trigger, hence, the signal associated with the hardware trigger is always composed of at least a couple of photons and never of a single photon. As a consequence the hardware trigger is not used for the later-on analysis to prevent a systematic bias.

\newpage
The software trigger only searches for signal events within the unflagged region. It compares the ADC channel values (see figure~\ref{fig:trace_flagged}) of four consecutive samples. The trigger condition is fulfilled when the sum of two samples subtracted from the sum of the following two samples is larger than 20 ADC channels. Later on the number of identified physical events $N_\text{phys}$ by the described software trigger is used for a correction of the signal efficiency.

The flagging of samples before each motor interference prevents that long physical events might start in the unflagged region and run into the motor interference. Such events have to be discarded. 
Since this can happen more often for long signal events, leading to a systematic bias of throwing away more longer events, the flagging is done in a way to prevent this.
The flagged region before each motor interference is 9\,$\mu$s long, separated into 8\,$\mu$s for the event reconstruction and 1\,$\mu$s for the baseline determination after an event. The flagged region after each motor interference is 1\,$\mu$s long and can be used for the baseline determination before an event.

The maximum length of the event window is set to 8\,$\mu$s, which corresponds to approximately 4.7 -- 8 lifetimes of the triplet state of LAr. This reduces the signal efficiency by 0.03 -- 0.9\% due to the rejection of events which are longer than 8\,$\mu$s. 

While the position of the baseline before an event is fixed, the position of the baseline after an event can be shifted by a maximum of 10\,$\mu$s in cases where the baseline is not sufficiently smooth directly after an event. This increases the fraction of accepted events for the further analysis.

Finally, the samples before and after the periodic interference from the stepper motor can be used for the event reconstruction and baseline determination, but a software trigger has to lie within the unflagged region (see figure~\ref{fig:trace_flagged}). 

\begin{figure}[ht!]
\centering
\includegraphics[width=8.8cm]{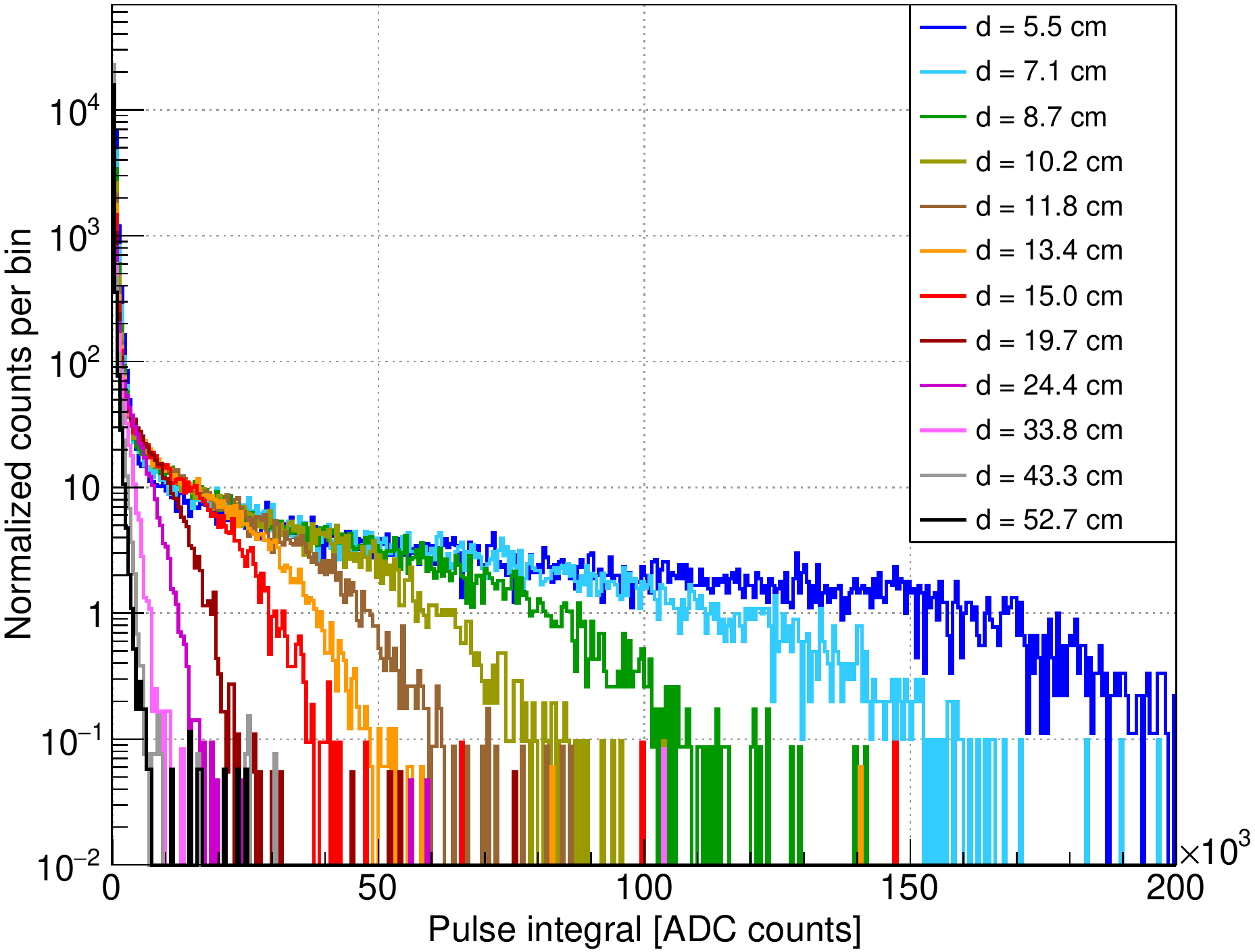}
\caption{(Color online) Pulse integral spectra of the \isotope[90]{Sr} source for various distances, normalized by the live time of the detector and the ratio of accepted vs. identified physical events.}
\label{fig:FADC_spectra}
\end{figure}

\newpage
Physical events starting in the unflagged area, which enter the area before the motor interference, but end at least 1\,$\mu$s before the motor interference, will be accepted for further analysis. On the contrary, physical events running into or starting within a motor interference might be identifiable, but will be discarded due to the missing post baseline.
If the variance of the baseline before and after the physical event is sufficiently small, the event is integrated over the signal window. 
After subtracting the baseline level, the integral enters a histogram as shown in figure~\ref{fig:FADC_spectra}.

The constructed histograms for each distance are divided by the live time of each individual measurement and corrected for the rejected events, respectively. 
The latter is achieved with a correction factor $c$, which is the number of accepted events $N_\text{acc}$ divided by the number of identified events $N_\text{phys}$ by the software trigger: $c = N_\text{acc} / N_\text{phys}$.

\begin{figure}[ht!]
\centering
\includegraphics[width=8.8cm]{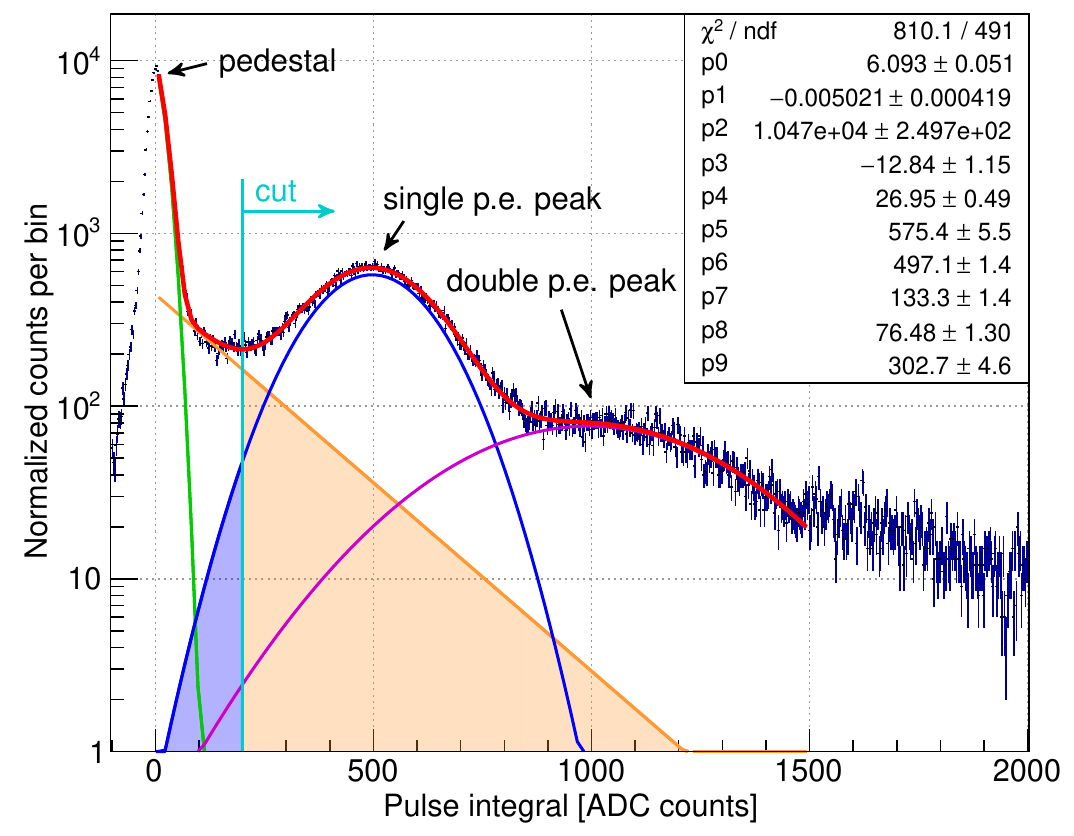}
\caption{(Color online) Example of a pulse integral spectrum with the combined fit (red) of the noise pedestal, single p.e.\ peak and double p.e.\ peak. The vertical line (light blue) indicates the cut at the minimum of the valley between pedestal and single p.e.\ peak. The spectrum is integrated on the right side of this cut. The exponential part (orange) of the pedestal on the right side of the cut is subtracted from the integral and the missing Gaussian part (dark blue) of the single p.e.\ peak on the left side is added to the integral.}
\label{fig:gausfit}
\end{figure}

From the spectra in figure \ref{fig:FADC_spectra}, which have to be calibrated and integrated, the average incident photon rate is determined. To correct for accidental noise triggers, which form the pedestal indicated in figure~\ref{fig:gausfit}, the low end of each pulse integral spectrum is fit by a combination of an exponential and three Gaussian functions describing the pedestal, the single and double photoelectron (p.e.) peak, which follows the description in \cite{combined_fit}.  
\begin{align}
f(x)
= \underbrace{ \mathrm e^{p_0+p_1 \cdot x}
+ p_2 \cdot \mathrm e^{-\frac{(x-p_3)^2}{2 \cdot p_4^2}}}_{\text{pedestal}}
&+ \underbrace{p_5 \cdot \mathrm e^{-\frac{(x-p_6)^2}{2 \cdot p_7^2}}}_{\text{single p.e. peak}}\\
&+ \underbrace{p_8 \cdot \mathrm e^{-\frac{(x-2 \cdot p_6)^2}{2 \cdot p_9^2}}}_{\text{double p.e. peak}}\nonumber
\end{align}

\newpage
A cut to eliminate the pedestal was applied to the local minimum between the pedestal and single p.e. peak, above which real photon signals dominate the dataset.
An example of such a fit is shown in figure~\ref{fig:gausfit}.

Finally, the average incident photon rate is calculated by integrating the histogram, weighted by the respective photoelectron value of each bin. This integral is corrected by subtracting the exponential part of the pedestal on the right side of the cut and adding the missing Gaussian part of the single p.e.\ peak on the left side. 
The corrected value is proportional to the number of photons that hit the PMT, i.e.\ the light intensity at the measured distance $d$. The light intensity $I$ is dependent on the initial intensity $I_0$ and follows the Beer-Lambert law.
\begin{align}
I(d) = I_0 \cdot \mathrm e^{-d / \alpha_\text{att}} \quad \text{with} \quad \frac{1}{\alpha_\text{att}} = \frac{1}{\alpha_\text{abs}} + \frac{1}{\alpha_\text{scat}}
\label{eq:attenuation}
\end{align}
From this the attenuation length $\alpha_\text{att}$, composed of absorption length $\alpha_\text{abs}$ and the scattering length $\alpha_\text{scat}$, can be derived.

Equation~\ref{eq:attenuation} does not include any background light source, but since the electrons emitted by the \isotope[90]{Sr} source can create Cherenkov light inside LAr an additional term must be added to account for it. The Cherenkov background will be determined by Monte Carlo simulation studies of the setup.

\newpage
\section{Simulation studies}

A Monte Carlo simulation of the setup was performed in order to model the experimental results and to test the analysis procedure on simulated data. 

The absorption length is implemented in a wavelength dependent way following reference \cite{att_meas_6} and scaled to achieve a certain absorption length at 128\,nm in the simulation. However, the measurement is is not sensitive to the wavelength since the scintillation spectrum is shifted by the WLS and accumulated by the PMT. Because of this it is not possible to verify the implemented absorption length per wavelength using the measured data.

Another drawback is that a wavelength resolved measurement could identify impurities like xenon and oxygen due to their characteristic emission peaks \cite{LAr_spectrum}. 
While small amounts of xenon in LAr can improve the overall light yield, it has been shown that xenon impurities lead to absorption bands in the region of the LAr scintillation peak, which can drastically decrease the absorption length \cite{att_meas_6}. None of these effects can be addressed with the performed measurement, hence the result is specific for the LAr setup in \textsc{Gerda}.

\subsection{Rayleigh scattering}
\label{sec:raygleigh_scatt}

Calculations and former measurements regarding Ray\-leigh scattering in LAr can be found in Ref. \cite{rayleigh_calculation, rayleigh_measurement}. However, there is a discrepancy between the obtained values at 128\,nm --- the calculation results in 90\,cm, while $66 \pm 3\,\unit{cm}$ were measured. 

A scattering length of \mbox{$\alpha_\text{scat}(128\,\unit{nm}) = 70\,\unit{cm}$} in LAr, which is closer to the quoted measurement, is implemented into the simulation framework for the attenuation measurement in \textsc{Gerda} following the description in \cite{rayleigh_calculation} and \cite{refractive_indices}. It should be noted, that $\alpha_\text{scat}$ strongly changes over the width (FWHM $\sim$ 6\,nm \cite{scintillation_peak}) of the 128\,nm emission peak of the scintillation light. In fact it is $\alpha_\text{scat}(122\,\unit{nm}) = 43\,\unit{cm}$ and increases to $\alpha_\text{scat}(134\,\unit{nm}) = 104\,\unit{cm}$, which is treated as a source of systematic uncertainty.

\subsection{Background consideration}
\label{sec:background_sources}

All relevant background sources have to be implemented into the simulation to model the measured data. A known background inside the LAr in \textsc{Gerda} is coming from \isotope[39]{Ar}, with an activity of roughly 1.4\,Bq/l \cite{argon_activity}. \isotope[39]{Ar} is a pure beta emitter with an endpoint energy of 565\,keV, which is equivalent to a CSDA (continuous slowing down approximation) range for electrons of 2\,mm in LAr and 0.4\,mm in steel\footnote{Since the material steel is not available in the ESTAR program, the values for iron, nickel and chromium were used for an estimation.}
\cite{estar}. The thickness of the steel housing of the setup is 2.6\,mm, hence, it is not expected that any \isotope[39]{Ar} outside of the setup can contribute to the scintillation light detection. The setup itself contains roughly 7\,l of LAr, resulting in about 10\,Bq of \isotope[39]{Ar} that can produce scintillation light inside the setup. 
Hamamatsu quotes an activity of 16\,mBq \cite{pmt_activity} for the operated PMT \mbox{R11065-10}, which is composed of  \isotope[226]{Ra}, \isotope[228]{Th} and dominated by \isotope[40]{K}.
Compared to the activity of the \isotope[90]{Sr} source of 7\,kBq the expected background from of \isotope[39]{Ar} and the PMT are negligible.

Another potential background can originate from the setup itself. It is made out of stainless steel (\href{https://en.wikipedia.org/wiki/SAE_304_stainless_steel}{SAE 304}). Measurements of the radiopurity of various kinds of steel are summarized in \cite{steel}. A conservative estimation does not exceed 3\,Bq/kg and taking into account the weight of the setup of roughly 13\,kg the background from steel can be neglected as well.

The radioactive source that is used to trigger the scintillation light is made of \isotope[90]{Sr} which decays into \isotope[90]{Y} as shown in figure~\ref{fig:Sr90_decay_chain}. The maximum electron energy from this decay chain is 2280\,keV. The minimum electron energy required to produce Cherenkov light inside LAr is about 260\,keV for the creation of optical photons, but only 150\,keV for photons of 115\,nm wavelength. The reason for this is the refractive index which grows with smaller wavelengths. The refractive index of LAr is implemented in the simulation based on the measurements in \cite{refractive_indices} and extrapolated down to 110\,nm.

Since Cherenkov light features a continuous spectrum over a wide wavelength range \cite{cherenkov} it has to be considered as a background for the attenuation length measurement. 
While the WLS shifts UV light to the optical region, it is transparent to optical photons. Additionally, photons in the optical region have a much longer range in LAr, since their energy is too low to excite argon atoms. Furthermore, the reflectivity of steel is different for the UV and optical range, as discussed in section~\ref{sec:reflectivity}, which affects the traveled path of optical photons in LAr since they are less often absorbed at the steel surface of the setup.
In conclusion, Cherenkov light causes a non-trivial and non-constant background for the various measuring points and has to be modeled with the simulation framework.

\subsection{Steel reflectivity}
\label{sec:reflectivity}

\begin{figure}[ht!]
\centering
\includegraphics[width=8.8cm]{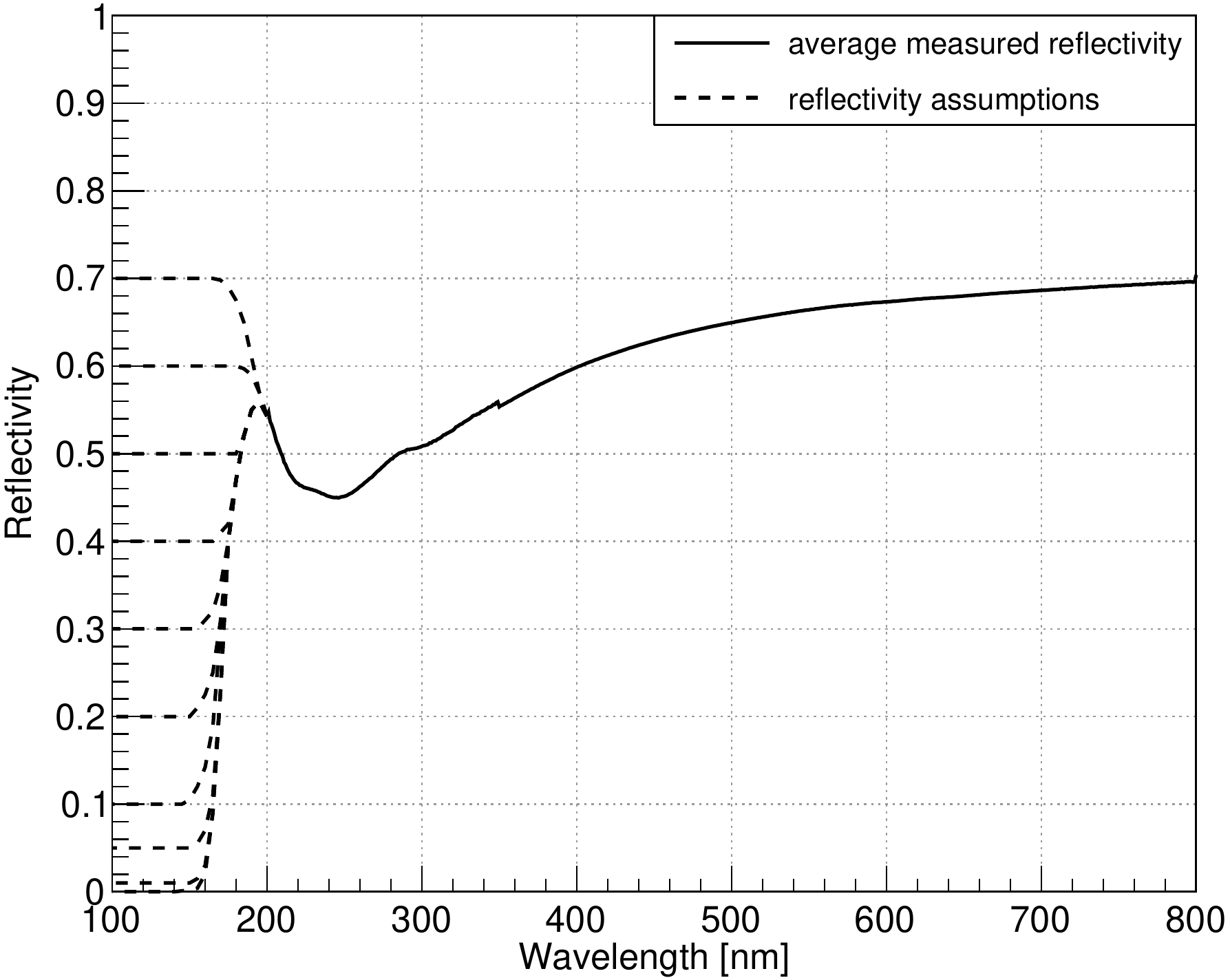}
\caption{Reflectivity measurement of the electropolished steel between 200\,nm and 800\,nm. The values below 200\,nm have to be assumed for the implementation in the simulation.}
\label{fig:reflectivity}
\end{figure}

Reflections at the electropolished steel surface of the setup have to be taken into account to determine an effective solid angle correction, which is especially important for the treatment of Cherenkov light. The reflectivity of the steel used for the setup has been measured between 200\,nm and 800\,nm in the Leibniz Institute of Polymer Research Dresden (IPF) with a \href{https://www.agilent.com/en/products/uv-vis-uv-vis-nir/uv-vis-uv-vis-nir-systems/cary-5000-uv-vis-nir}{Cary 5000} spectrophotometer. The device is not able to measure at lower wavelengths, which is why the reflectivity has to be estimated below 200\,nm for the simulation. The measured values as well as the assumptions for the simulation are shown in figure~\ref{fig:reflectivity}. Which reflectivity assumption is used in the simulation is indicated by the reflectivity at 128\,nm in the following.

\newpage
\subsection{Solid angle correction and Cherenkov background}
\label{sec:solid_angle}

To determine the attenuation length from equation~\ref{eq:attenuation}, a solid angle correction has to be applied to the light intensity measured at each distance. Monte Carlo simulations were performed to investigate the solid angle correction which is influenced by several effects making it non-trivial. 

First of all, the electrons from the \isotope[90]{Sr} source can travel a few millimeters in the LAr and produce scintillation photons on their path. Additionally, it is also possible that electrons create bremsstrahlung photons which trigger further electrons via Compton scattering or photoeffect farther away from the \isotope[90]{Sr} source. This leads to an extended virtual source of scintillation photons that cannot be described as point-like anymore.

In addition, wavelength shifted photons are emitted isotropically and can reach the PMT via reflections or scatterings from any location within the setup. Since such photons are in the optical region, where the reflectivity of the steel is high (fig.~\ref{fig:reflectivity}), they can travel a very long distance in LAr increasing the probability for triggering the PMT. 

Furthermore, photons can be reflected on the steel of the setup or Rayleigh scatter inside the LAr which both elongates their traveled distance and increases the distance uncertainty. This is especially important for Cherenkov photons, since the reflectivity is much higher in the optical region (fig.~\ref{fig:reflectivity}), where the PMT is more sensitive (fig.~\ref{fig:PMT_eff+TPB}).

\begin{figure}[ht!]
\centering
\includegraphics[width=8.8cm]{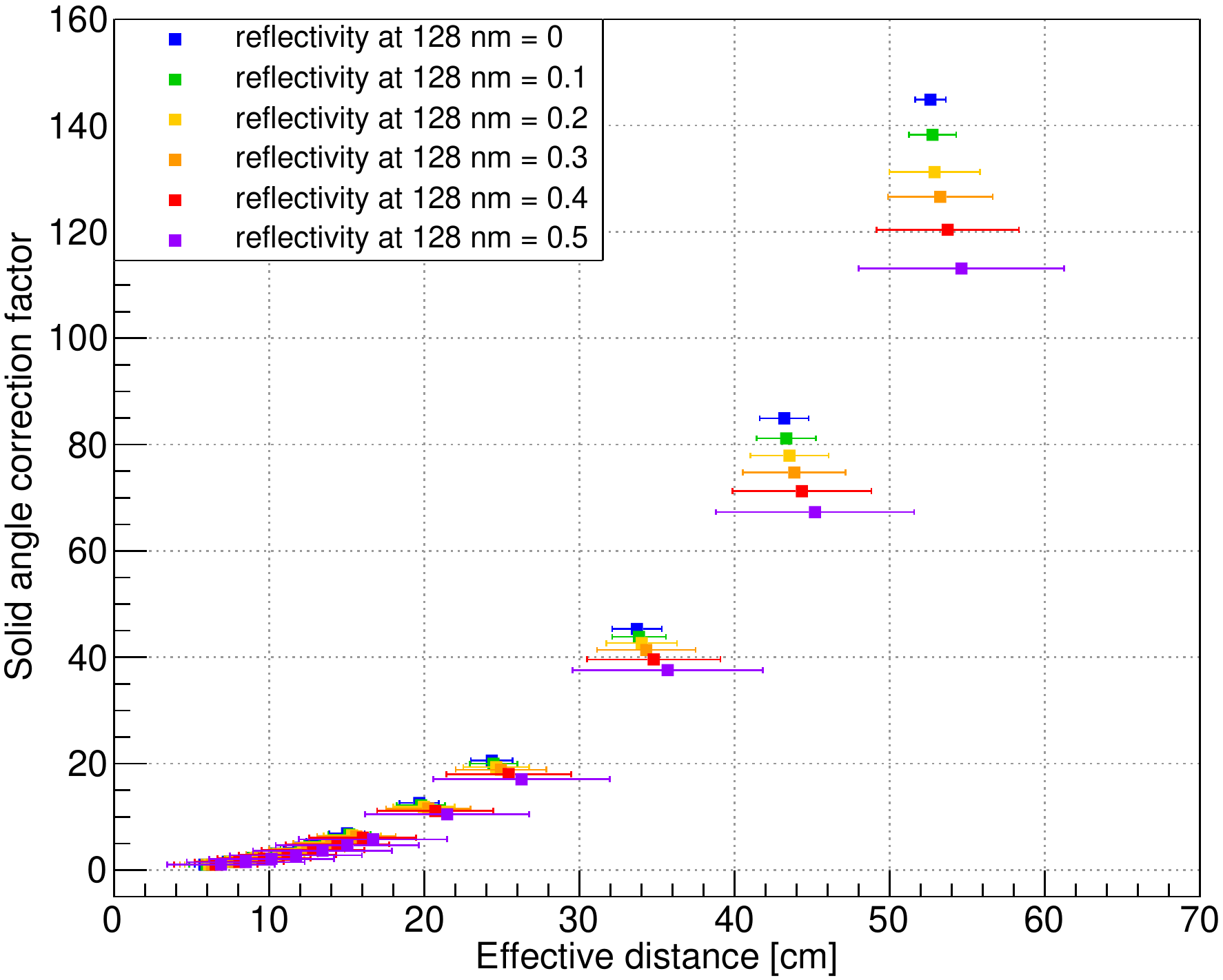}
\caption{Dependency of the solid angle correction factor on the reflectivity at 128\,nm using six different assumptions. The simulation uses an absorption length of 1\,km, a scattering length of 70\,cm and an effective light yield of 5000\,$\gamma$/MeV. The uncertainty on the distance grows with the reflectivity because more photons can travel a longer path in LAr due to the increased probability to get reflected at the steel surface of the setup.}
\label{fig:solid_angle_refl}
\end{figure}

With the help of detailed Monte Carlo simulations it was found that the solid angle correction depends on the reflectivity assumption at 128\,nm as shown in figure~\ref{fig:solid_angle_refl}. 

\begin{figure}[ht!]
\centering
\includegraphics[width=8.8cm]{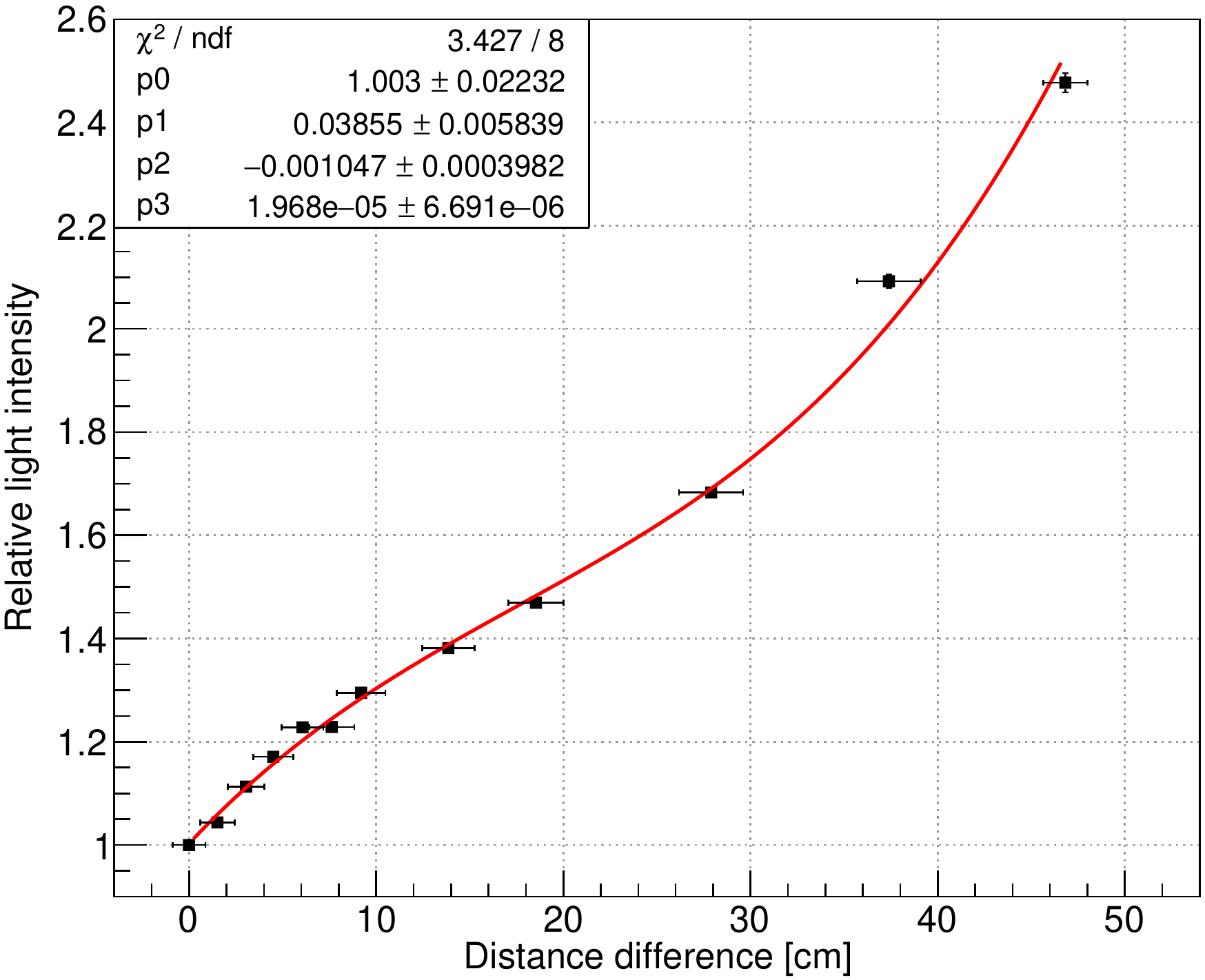}
\caption{Polynomial fit of the Cherenkov background after solid angle correction for a reflectivity of 0\% at 128\,nm. The number of Cherenkov hits of the PMT are normalized by the first measuring point and plotted over the distance difference $\Delta d$ to the first point. Due to the high reflectivity of the steel in the optical region the Cherenkov background increases with the distance.}
\label{fig:cherenkov_bkg_fit}
\end{figure}

Since it is not possible to distinguish in the experimental data which PMT hits were caused by scintillation photons or Che\-renkov photons, all PMT hits will be corrected for the respective solid angle. 
Additionally, it was found that the Cherenkov background depends on the reflectivity assumptions in the simulation.
In order to be able to include the Cherenkov background in equation~\ref{eq:attenuation} for a combined fit of signal and background, an analytical description is required. 
The following fit formula (eq.~\ref{eg:cherenkov_bkg_fit}) was found to describe the Cherenkov background well for any reflectivity assumption. 
Figure~\ref{fig:cherenkov_bkg_fit} shows an example of the Cherenkov background fit with equation~\ref{eg:cherenkov_bkg_fit}. 
\begin{align}
\label{eg:cherenkov_bkg_fit}
b(\Delta d) = p_0 + p_1 \cdot \Delta d + p_2 \cdot (\Delta d)^2 + p_3 \cdot (\Delta d)^3
\end{align}
The polynomial describes the Cherenkov background $b$ with respect to the first measuring point whereas the distance difference between the measuring points is indicated by $\Delta d$.

\subsection{Simulation results}
\label{sec:analysis_sim}

To probe the anticipated analysis a series of Monte Carlo simulations was generated using different sets of input parameter values. 
For this the absorption length was varied within $10-30\,\unit{cm}$, the effective LAr light yield between $1000 - 10000\,\unit{\gamma/MeV}$ and the reflectivity at 128\,nm was modified as shown in figure~\ref{fig:reflectivity}. These simulations were then analyzed using the same algorithms as for the real data aiming to recover the original simulation input values.

\begin{figure}[ht!]
\centering
\includegraphics[width=8.8cm]{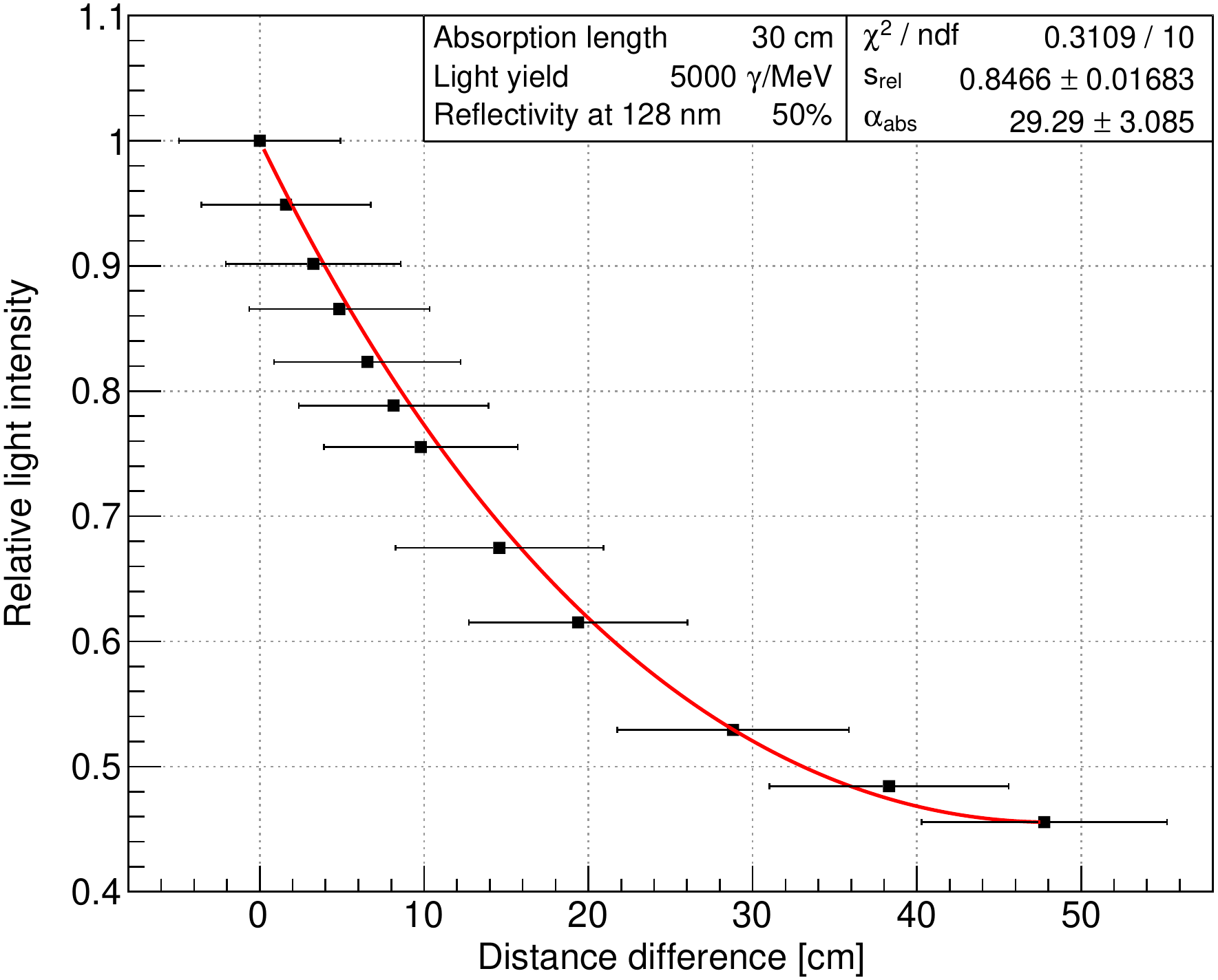}
\vspace{0.2cm}
\includegraphics[width=8.8cm]{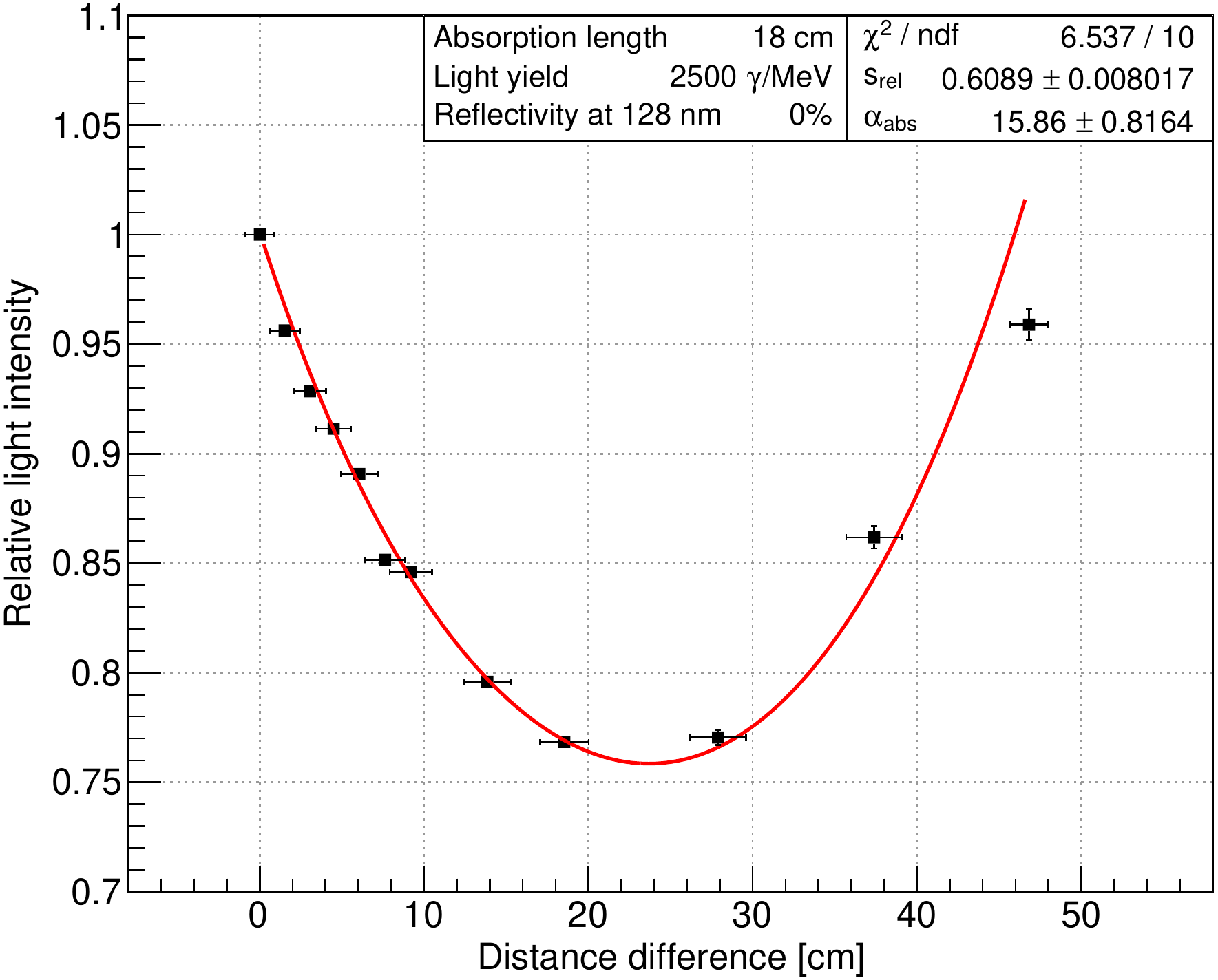}
\caption{Number of scintillation photons detected by the PMT relative to the first point and combined fit of signal and background for two simulations with various input parameters. 
In the upper plot no influence of Cherenkov light is visible at all while in the lower plot Cherenkov light is dominating at the larger distances and the relative light intensity grows, which is mainly due to the different values of the effective light yield.
With a higher reflectivity (top) photons get more often reflected at the steel surface, which leads on average to a farther traveled distance and is increasing its uncertainty.}
\label{fig:simulation_combined_fit}
\end{figure}

The combined fit $f(\Delta d)$ includes the signal $s$ coming from the scintillation photons and the background $b$ from the Cherenkov light. The fit of the Cherenkov background (eq.~\ref{eg:cherenkov_bkg_fit}) is performed beforehand and enters the combined fit with its parameters fixed. Equation~\ref{eq:combined_fit} shows the combined fit formula and figure~\ref{fig:simulation_combined_fit} the application to two examples of the simulation using:
\begin{alignat}{2}
\label{eq:combined_fit}
f(\Delta d) &= s_\text{rel} \cdot s(\Delta d) + (1 - s_\text{rel}) \cdot b(\Delta d) \\
s(\Delta d) &= \mathrm e^{-\Delta d / \alpha_\text{abs} }.
\end{alignat}
The fit parameter $s_\text{rel}$ reveals the signal strength relative to the sum of signal and background. It gives an indication of the effective LAr light yield, hence a high value means that the light yield, i.e.\ the number of produced scintillation photons is high. Physical results for the parameter $s_\text{rel}$ have to be between 0 and 1. The parameter $\alpha_\text{abs}$ gives the absorption length in cm. 

Figure~\ref{fig:simulation_combined_fit} also shows that if the absorption length is very short and the effective light yield is very low (bottom), the influence of the Cherenkov effect is very well visible for larger distances, where less scintillation light reaches the PMT. This causes the increase of the relative light intensity above a distance difference of about 20\,cm. On the contrary, for longer absorption lengths and higher effective light yields (top) the impact of Cherenkov light is not seen at all and the relative light intensity keeps decreasing with growing distance difference.

\section{Data analysis}
\label{sec:analysis_data}

In this section the analysis technique successfully evaluated for the simulation in section~\ref{sec:analysis_sim} is applied to the measured data.
For each reflectivity assumption at 128\,nm a respective solid angle correction is obtained. Following, the data are analyzed with each solid angle correction and the combined fit $f(\Delta d)$ is applied according to equation~\ref{eq:combined_fit} using the Cherenkov background fit of the simulation. 

\newpage
\subsection{Determination of the best matching simulation input parameters}

With the help of the two fit parameters $s_\text{rel}$ and $\alpha_\text{abs}$ from equation~\ref{eq:combined_fit} the simulation that matches the data best is searched for using a $\chi^2$-method. 
By changing the absorption length between 10 -- 30\,cm and the effective light yield between 1000 -- 10000\,$\unit{\gamma/MeV}$ in an iterative way, the best matching simulation input parameters, i.e., absorption length and effective light yield, are determined for each reflectivity assumption.

\newpage
Since the reflectivity of steel at 128\,nm is unknown and the solid angle correction depends on it, the effect on the fit parameters $s_\text{rel}$ and $\alpha_\text{abs}$ is studied by applying a solid angle correction based on a certain reflectivity on a simulation using a different reflectivity assumption.
For this purpose the respective best matching simulations for each reflectivity assumption are analyzed using the available solid angle correction for each reflectivity assumption.
 
\begin{figure*}[ht!]
\centering
\includegraphics[width=7.5cm]{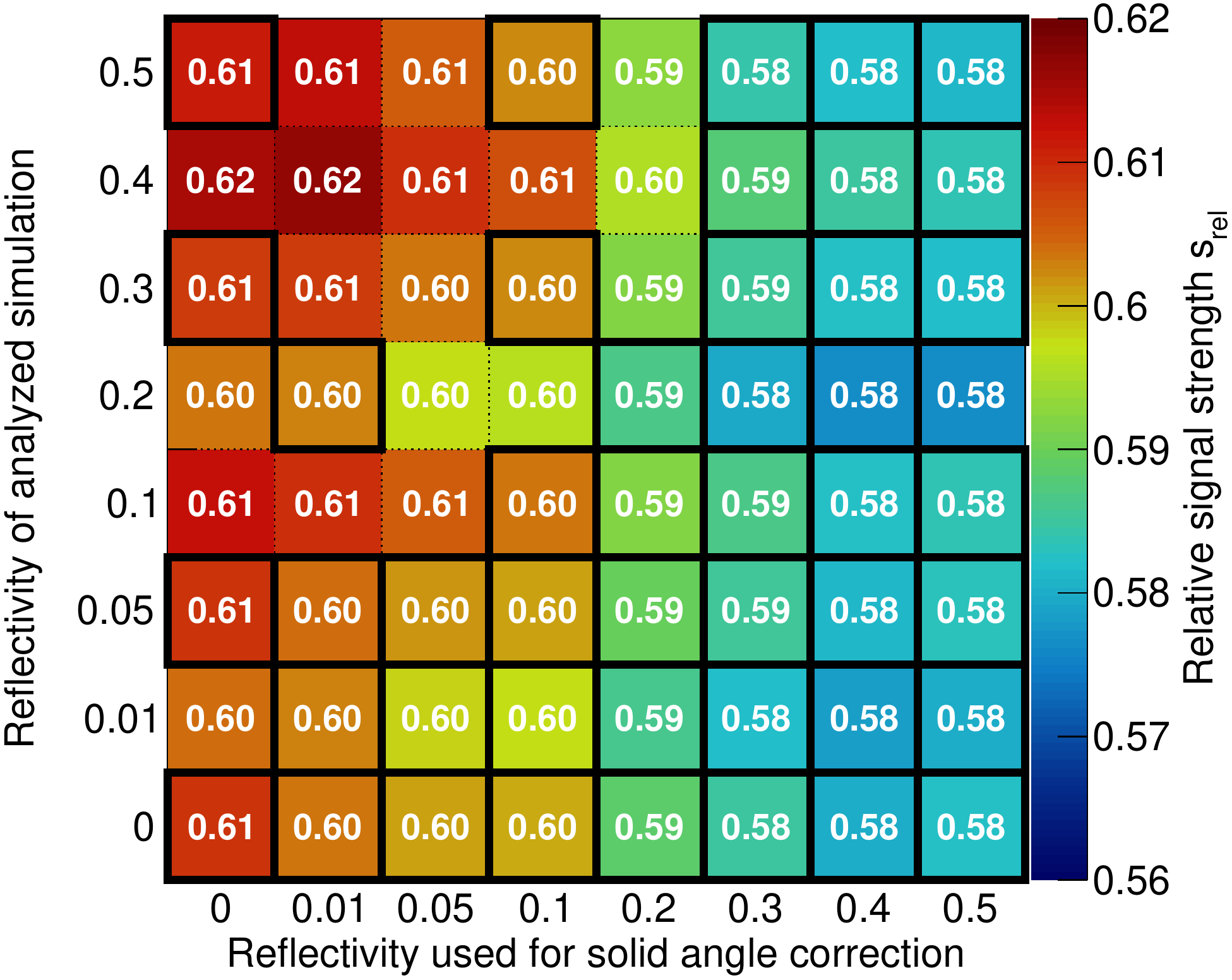}
\qquad
\includegraphics[width=7.5cm]{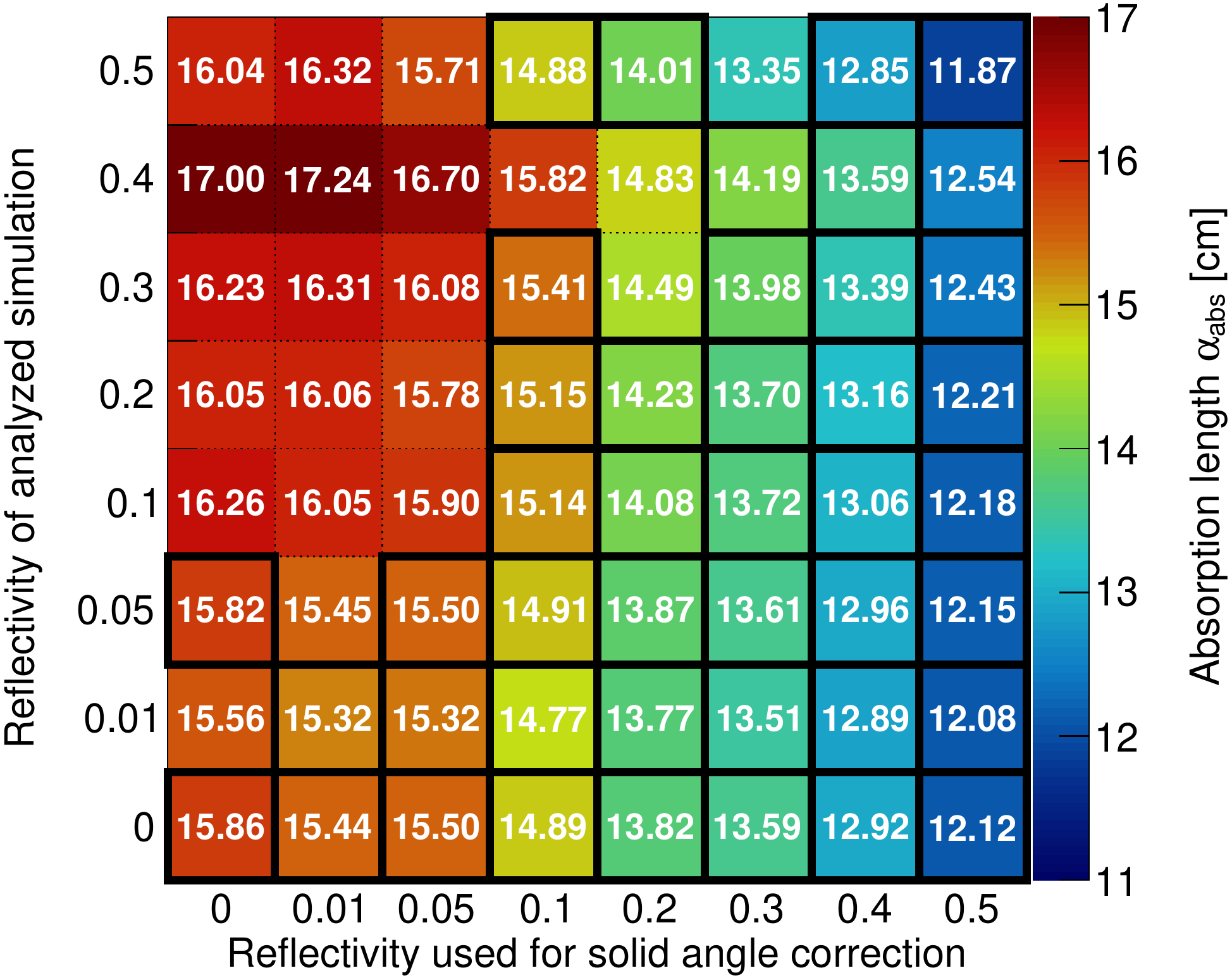}
\caption{Values of the fit parameters $s_\text{rel}$ (left) and $\alpha_\text{abs}$ (right) for the best matching simulations of each reflectivity assumption analyzed with solid angle corrections of all reflectivity assumptions. 
The black boxes indicate the values of $s_\text{rel}$ and $\alpha_\text{abs}$ that match with the experimentally determined fit values within their uncertainties, obtained from an optimization procedure. 
If a simulation of fixed parameter space (plotted on the left axis) matches with the experimental data to the demanded accuracy, there is a horizontal line of highlighted bins (black frames). This indicates that the simulation behaves like the experimental data when analyzed with all reflectivity assumptions (plotted on the bottom axis).
While individually considered a reflectivity below $5\%$ is favored, both plots together agree to a reflectivity of $0\%$ at 128\,nm as best match.  
}
\label{fig:fit_comparison_2d}
\end{figure*}

Figure~\ref{fig:fit_comparison_2d} shows the values for the fit parameters $s_\text{rel}$ and $\alpha_\text{abs}$ for all cases of simulations analyzed with different solid angle corrections.

The relative signal strength $s_\text{rel}$ is correlated to the effective light yield in the simulation and decreases slightly while the reflectivity of the applied solid angle correction is increased. 
The extracted absorption length $\alpha_\text{abs}$ tends to underestimate the input value of the generated simulation when analyzing it with a solid angle correction using a low reflectivity.
The fit value of $\alpha_\text{abs}$ converges to the input value for higher effective light yields and higher reflectivities, since the influence of the Cherenkov background decreases in those cases.

\subsection{Comparison of simulation and experimental data}
\label{sec:comparison_sim_data}

The experimental data are also analyzed using the solid angle correction for each reflectivity assumption. The black boxes in figure~\ref{fig:fit_comparison_2d} indicate the values of $s_\text{rel}$ and $\alpha_\text{abs}$ of the simulations that are within the uncertainties of the fit parameters analyzed with the respective solid angle correction. The goal is to find the simulation that describes the experimental data the most accurately when analyzing it with the different solid angle corrections arising from the underlying reflectivity assumptions.

As figure~\ref{fig:solid_angle_refl} already indicates, the uncertainty on the distance grows with higher reflectivity at 128\,nm, because photons reflected more often can reach the PMT, which increases their average traveled path and also leads to a higher uncertainty. This increase in the uncertainty propagates onto the fit parameters $s_\text{rel}$ and $\alpha_\text{abs}$, whereas the effect is stronger on $\alpha_\text{abs}$. This is why all simulations are highlighted as matching to the experimental data for high reflectivity assumptions used for the solid angle correction in figure~\ref{fig:fit_comparison_2d}. 

From the plots in figure~\ref{fig:fit_comparison_2d} it can be seen that the relative signal strength $s_\text{rel}$ favors the reflectivity assumptions of 0\% and 5\% and the absorption length $\alpha_\text{abs}$ the reflectivity models at 0\%, which is in agreement with each other. 
The best matching simulation for a reflectivity of 0\% at 128\,nm is generated with an absorption length of 18\,cm and an effective light yield of 2500\,$\gamma$/MeV.

The fit results of experimental data and best matching simulation agree very well with each other, which is shown in figure 12, although the measured data points feature a wider spread.

One reason for the larger spread of the experimental data could be that the stepper motor adjusting the distance between PMT and source has been characterized at room temperature. 
The according step width is expected to change while operating the setup in LAr due to the low temperature.
The shrinkage of the steel setup is about 0.3\% for 88\,K \cite{steel_coeff}, which reduces the distance by 1.6\,mm for the farthest measuring point. 
In addition the uncertainty inferred from the step width of the motor has been determined by moving the source several times to the maximum distance while operating the setup in LAr. The average uncertainty of the source movement is 1.3\,mm.
In combination with the absolute distance uncertainty of 1\,mm from the measurement of the shortest distance the total uncertainty for the closest measuring point is 2.4\,mm and the for the farthest 3.9\,mm.

\subsection{Results and discussion}

The attenuation measurement in the \textsc{Gerda} LAr cryostat results in an absorption length of 
\begin{align}
\label{eq:result}
\alpha_\text{abs} = 15.8 \pm 0.7 (\text{stat}) {}^{+1.5}_{-3.2} (\text{syst})\,\unit{cm}
\end{align}
with the assumption of $\alpha_\text{scat}(128\,nm) = 70\,$cm and a steel reflectivity of 0\% at 128\,nm. 
 
The systematic uncertainty is obtained by analyzing the measured data with various solid angle corrections that are each based on simulations with a fixed set of simulation input parameters. Afterwards the difference of the results to equation~\ref{eq:result} are calculated. 

The following parameters were modified with respect to their default values: the steel reflectivity at 128\,nm was increased from 0\% to 5\%, the effective area of the PMT was changed from the minimum to the maximum physical diameter, $\alpha_\text{scat} (128\,\unit{nm})$ was set to the values derived from the width of the scintillation peak as described in section \ref{sec:raygleigh_scatt} and the distance between PMT and source was changed according to the total uncertainty derived in section \ref{sec:comparison_sim_data}.
The investigated systematics with the respective difference to the resulting absorption length are listed in table \ref{tbl:syst_uncert}.

An additional systematic uncertainty is created by the spectra sampling of the data due to the much lower statistics for the farther distances. 
The effect was studied by performing several simulations with the best matching input parameters and similar statistics as in the experimental data.
Following, the standard deviation of the difference with respect to equation~\ref{eq:result} is used as a systematic uncertainty as listed in table~\ref{tbl:syst_uncert}.

The attenuation length in the LAr in \textsc{Gerda} assuming $\alpha_\text{scat} = 70 \,cm$ at 128\,nm in derived according to equation~\ref{eq:attenuation}.
\begin{align}
\alpha_\text{att} = 12.9 \pm 0.5 (\text{stat}) {}^{+1.0}_{-2.3} (\text{syst})\,\unit{cm}
\end{align}

\begin{table}[ht!]
\centering
\renewcommand{\arraystretch}{1.2}
\label{tbl:syst_uncert}
\caption{Simulation input parameters for the investigation of the systematic uncertainties and the resulting difference to the absorption length quoted in equation \ref{eq:result}.}
\vspace{0.1cm}
\begin{tabular}{cr}
\toprule
simulation input parameter & syst. uncertainty \\
\midrule
steel reflectivity at 128\,nm & $-0.4$\,cm \\
efficient area of the PMT & $-2.7$\,cm \\
scattering length at 128\,nm & ${}^{+0.2}_{-0.5}$\,cm \\
distance between PMT and source & ${}^{+1.4}_{-1.4}$\,cm \\
statistics of the spectra sampling & ${}^{+0.3}_{-0.3}$\,cm \\
\bottomrule
\end{tabular}
\end{table}

\begin{figure}[ht!]
\centering
\includegraphics[width=8.8cm]{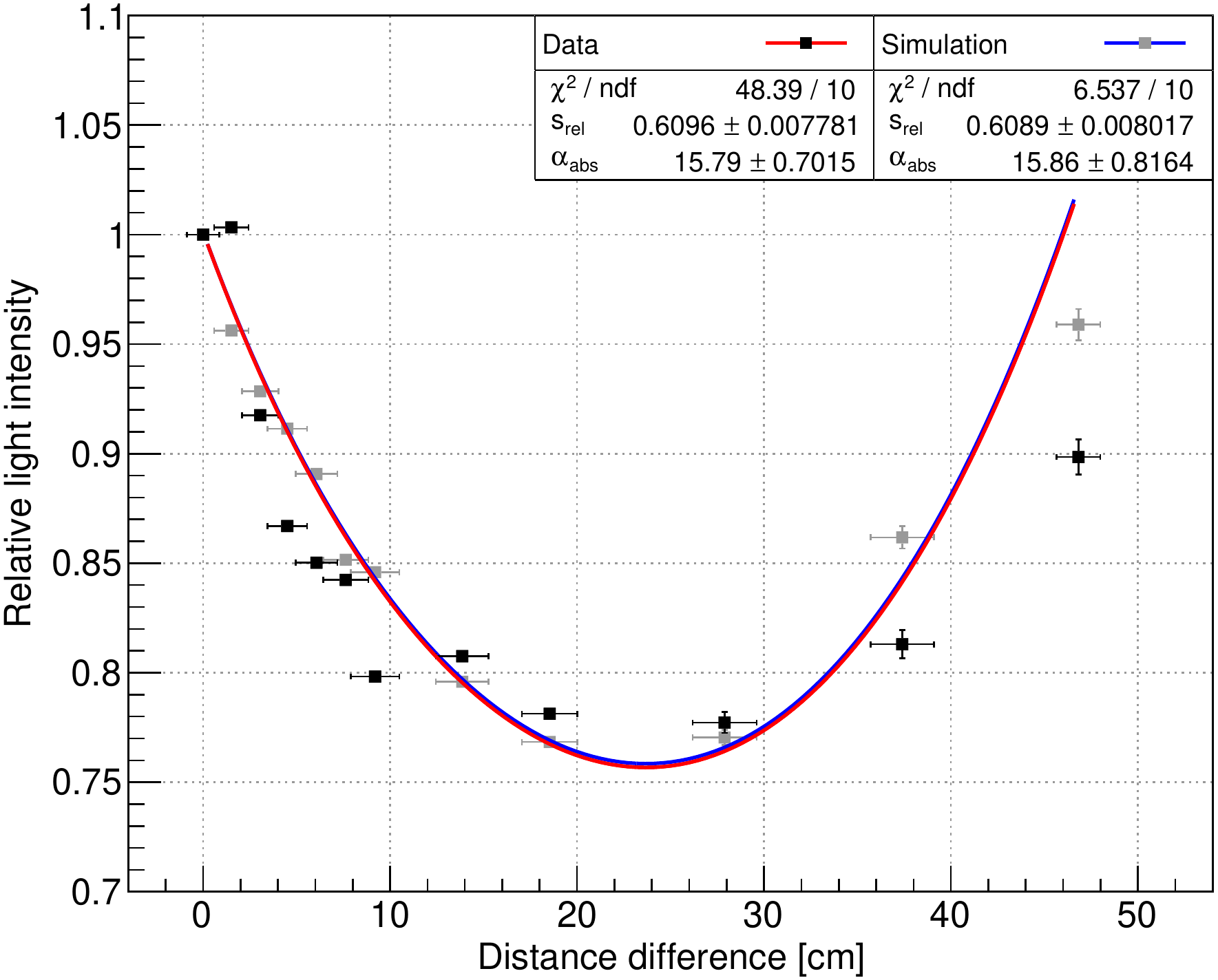}
\caption{Comparison of data and best matching simulation analyzed with the solid angle correction of a reflectivity assumption of 0\% at 128\,nm. The simulation was generated with an absorption length of 18\,cm, an effective light yield of 2500\,$\gamma$/MeV and a reflectivity of 0\% at 128\,nm.}
\label{fig:fit_comparison_data_sim}
\end{figure}

The major contribution to the uncertainty on the fit result of $\alpha_\text{abs}$ as shown in figure~\ref{fig:fit_comparison_data_sim} is coming from the uncertainty on the effective distance between source and PMT (relative uncertainty: 10.6\% for the closest distance, 1.9\% for the farthest distance). This is a direct consequence of the average path length a scintillation photon travels from its point of origin until reaching the WLS on the PMT. 
Another contribution to the uncertainty arises from the Monte Carlo statistics of the solid angle correction, which gets poorer for the measuring points that are farther away from the PMT (relative uncertainty: 0.1\% for the closest distance, 0.8\% for the farthest distance).

\section{Conclusion and Outlook}
\label{sec:conclusions}

An attenuation length measurement was performed in the \textsc{Gerda} LAr cryostat with a dedicated setup. 
The attenuation length has been extracted from the experimental data using detailed simulations of the setup to take into account Cherenkov background, geometrical acceptance of the PMT as well as details of the light propagation.
The simulation framework was further used to cross-check the subsequent analysis before applying it to the measured data.
The input parameters which lead to the best matching simulation are an absorption length of 18\,cm, a scattering length of $70\,\unit{cm}$ at 128\,nm, an effective light yield of 2500\,$\gamma$/MeV and a reflectivity of the steel of the setup of 0\% at 128\,nm. The best matching values of these parameters are compatible with the data.

It should be noted that the effective light yield is a combination of several efficiencies including the PMT efficiency, the WLS yield as well as the WLS absorption and emission spectrum. Hence, the effective light yield quoted here is specific to the setup and should not be compared to other measurements.

Since the measurement is not wavelength resolved, impurities like xenon or oxygen cannot be identified and the result reflects the absorption length of LAr in \textsc{Gerda} according to the impurity content at the time of the measurement.

In comparison to previous studies \cite{att_meas_1, att_meas_2, att_meas_3} the determined value of $\alpha_\text{att}$ turns out to be significantly smaller, but lies at the same order of magnitude (e.g. ref. \cite{att_meas_1}, 50\,cm).
However, in previous measurements very pure LAr was taken initially and doped with impurities in order to measure the attenuation dependency on the impurity composition. 
The studies using the purest LAr samples in an attenuation length of 1.10\,m \cite{att_meas_2} and $30-1790$\,m \cite{att_meas_3}.
On the contrary the LAr in the GERDA cryostat has not been purified and dwells in the cryostat since December 2009.

Despite the fact that the attenuation length and the effective light yield in this measurement are comparatively low, the LAr veto of\textsc{Gerda} performs well, because all parts are coated with WLS, except for the germanium detectors and their holders.  This ensures that a high fraction of the original scintillation light is directly shifted to the optical region maximizing the probability to trigger the veto's readout.

After \textsc{Gerda} finishes Phase II and the detector array is retrieved from the LAr cryostat, the attenuation measurement could be redone to check whether the LAr properties changed. It is especially interesting to investigate the LAr quality over time since the successor experiment \textsc{Legend} \cite{legend} will be built in the same infrastructure.  

In principle the developed setup can also be used in additional experiments to measure noble gas properties like absorption length and triplet lifetime. In fact, the PMT and the source can be exchanged to cover other demands, where the used combination of a high-energy beta source, TPB as WLS and the PMT designed for cryogenic temperature operation are not optimal. 

The setup could be improved by constructing a mechanical holder that fixes the source in its position. In this way the interference from the motor occurring in the pulse traces could be completely extinguished.
In addition, an alpha source could be used eliminating any source related Cherenkov light contribution.
This would greatly simplify the analysis and improve the precision of the attenuation measurement.

\section*{Acknowledgments}

This work was supported in part by grants from BMBF, DFG, INFN, MPG, NCN, RFBR and SNF. The authors thank the workshop of TU Dresden for building the setup and the colleagues and the workshop of the MPIK in Heidelberg for the opportunity to test the setup in liquid nitrogen.
We thank in particular Bernhard Schwingenheuer for his continuous support and help during the preparation and execution of the measurement as well as the analysis period.
Additional thanks go to Anja Maria Steiner from the IPF in Dresden who made it possible to measure the reflectivity of the steel.
Furthermore we thank the Center for Information Services and High Performance Computing (ZIH) at TU Dresden for the generous allocation of computing power.
The \textsc{Gerda} collaboration thanks the director and the staff of the LNGS for their continuous strong support of the \textsc{Gerda} experiment.

\bibliographystyle{elsarticle-num}

\end{document}